\documentclass[]{IEEEtran}

\usepackage{tikz}
\usepackage[sumlimits,intlimits]{amsmath}
\usepackage{amsmath,amsfonts,amssymb}%
\usepackage{bm}%
\usepackage{graphicx,graphics}%
\usepackage{cases}%
\usepackage[noadjust]{cite}%
\usepackage{color}%
\usepackage{cite,url}
\usepackage{verbatim}
\usepackage{algorithm}
\usepackage{balance}
\usepackage{multirow}
\usepackage{stfloats}
\usepackage{subfig}
\usepackage{xspace}

\begin{document}

\pagestyle{empty}

\title{Training-Based Synchronization and Channel
Estimation in AF Two-Way Relaying Networks}

\author{\vspace*{-2pt}\hspace*{-8pt}
\authorblockN{Ali A. Nasir$^{\dag}$, Hani Mehrpouyan$^{\ddag}$, Salman Durrani$^{\dag}$, Steven D. Blostein$^{\S}$, and Rodney A. Kennedy$^{\dag}$,
}\\
{\small
$^{\dag}$ Research School of Engineering, The Australian National University, Australia.\vspace{0pt}\\
$^{\ddag}$ Department of Electrical and Computer Engineering and Computer Science, California State University, Bakersfield, USA.\vspace{0pt}\\
$^{\S}$ Department of Electrical and Computer Engineering, Queen's University, Kingston, Canada.\vspace{0pt}\\
email: ali.nasir@anu.edu.au, hani.mehr@ieee.org, salman.durrani@anu.edu.au, steven.blostein@queensu.ca, rodney.kennedy@anu.edu.au
}\vspace{-20pt}}

\maketitle

{\let\thefootnote\relax\footnotetext{ This research was supported under Australian Research Council's Discovery Projects funding scheme (project number DP140101133). \vspace{-00pt}}}

\begin{abstract}

Two-way relaying networks (TWRNs) allow for more bandwidth efficient use of the available spectrum since they allow for simultaneous information exchange between two users with the assistance of an intermediate relay node. However, due to superposition of signals at the relay node, the received signal at the user terminals is affected by \emph{multiple impairments}, i.e., channel gains, timing offsets, and carrier frequency offsets, that need to be jointly estimated and compensated. This paper presents a training-based system model for amplify-and-forward (AF) TWRNs in the presence of multiple impairments and proposes maximum likelihood and differential evolution based algorithms for joint estimation of these impairments. The Cram\'{e}r-Rao lower bounds (CRLBs) for the joint estimation of multiple impairments are derived. A minimum mean-square error based receiver is then proposed to compensate the effect of multiple impairments and decode each user's signal. Simulation results show that the performance of the proposed estimators is very close to the derived CRLBs at moderate-to-high signal-to-noise-ratios. It is also shown that the bit-error rate performance of the overall AF TWRN is close to a TWRN that is based on assumption of perfect knowledge of the synchronization parameters.

\end{abstract}

\section{Introduction}

Relaying is a key technology to assist in the communication between two user terminals, especially when there are large transmission distances between them \cite{Laneman-A-03}. Unidirectional (one-way) relaying supports communication from a source user to a destination user and has been widely studied in the literature \cite{Laneman-A-03}. On the other hand, in two-way relaying networks (TWRNs), the flow of information is bidirectional and the two users exchange information simultaneously with the assistance of an intermediate relay node \cite{Oechtering-08-A}. Thus, compared with one-way half-duplex relaying, bidirectional relaying is a spectrally more efficient relaying protocol \cite{Oechtering-08-A}. Both amplify-and-forward (AF) and decode-and-forward (DF) protocols have been developed for TWRNs. In contrast to the DF protocol, the AF protocol is widely adopted, as it requires minimal processing at the relay node \cite{Cheng-12-A}.

During the two phase communication in AF TWRNs, the two users first transmit their information to the relay node. The relay broadcasts its received signal to both users in the second phase.\footnote{To ensure spectral efficiency, three or four phase communication protocols for AF TWRN \cite{Tian-11-P} are not considered in this paper.} However, the two users' signals at the relay node undergo different propagation paths and may not be aligned in time and frequency. Consequently, the superimposed signal broadcasted from the relay node is affected by \emph{multiple impairments}, e.g., channel gains, timing offsets, and carrier frequency offsets. The existing literature does not take all these impairments into account in studying the performance of AF TWRNs \cite{Tian-11-P}. Although estimation and compensation algorithms have been proposed to counter these impairments in unidirectional relaying networks \cite{Tian-10-A,Nasir-13-A-TSP}, the proposed algorithms cannot be directly applied to TWRNs due to differences between the two system models. Particularly, in TWRNs (see Fig. \ref{fig:BD}), each user can exploit the knowledge of the self transmitted signal during Phase 1 in order to detect the signal from the other user during Phase 2. Recently, blind \cite{Abdallah-12-A} and semi-blind \cite{Abdallah-13-A} methods have been proposed for channel estimation only in AF TWRNs. A particle filtering based method for estimating channel and timing offset is proposed in \cite{Jiang-13-A}. For training-based methods, which are more suited for practical implementation \cite{Li-10-A}, channel estimation \cite{Wang-12-A,Gao-09-A,Cheng-12-A} or joint channel and carrier frequency offset estimation \cite{Wang-10-P} has been considered in the literature. However, to the best of authors' knowledge, an estimation and decoding scheme for TWRNs in the presence of channel gains, timing offsets, and carrier frequency offsets is still an open research problem.

In this paper, a complete synchronization approach, i.e., joint estimation and compensation of channel gains, timing offsets, and carrier frequency offsets for AF TWRNs is proposed. Upon reception of the superimposed signals broadcasted from the relay node, the user nodes first jointly estimate the impairments using \emph{known} training signals and the proposed maximum likelihood (ML) or differential evolution (DE) based estimators~\cite{Price-05-B}. Subsequently, the users employ the proposed minimum mean-square error (MMSE) receiver in combination with the estimated impairments to decode the received signal. Each user uses knowledge of its transmitted data to cancel out the self interference and decode the opposing user's signal. The contributions of this paper can be summarized as follows:

\begin{itemize}
\item We propose a system model for achieving synchronization and obtaining the channel parameters in AF TWRN.

\item We derive Cram\'{e}r-Rao lower bounds (CRLBs) for joint estimation of multiple impairments at the user nodes for TWRN. These bounds can be applied to assess the performance of synchronization and channel estimation in AF TWRN networks.

\item We derive an ML based estimator for joint estimation of multiple impairments. A DE based algorithm is proposed as an alternative to the ML estimator to significantly reduce the computational complexity associated with synchronization in AF TWRNs. Simulation results show that the mean square error (MSE) performances of both ML and DE estimators are close to the CRLB at moderate-to-high signal-to-noise-ratios (SNRs).

\item We derive an MMSE receiver for compensating the impairments and detecting the signal from the opposing user.

\item Extensive simulations are carried out to investigate the estimated MSE and bit-error rate (BER) performances of the proposed transceiver structure. These results show that the BER performance of an AF TWRN can be significantly improved in the presence of practical synchronization errors. In fact, the application of the proposed transceiver results in an overall network performance that is very close to that of an ideal network based on the assumption of perfect knowledge of synchronization and channel parameters.


\end{itemize}

The remainder of the paper is organized as follows: Section \ref{sec:SM} presents the system model while the CRLBs for the joint estimation problem are derived in Section \ref{sec:CRLB}. In Section \ref{sec:EST}, the ML and DE based estimators for joint estimation of multiple impairments at the user nodes are presented. In Section \ref{sec:SD}, the proposed MMSE receiver is derived, while Section \ref{sec:SR} presents the simulation results.

Notation: Superscripts $(\cdot)^{T}$, $(\cdot)^{\ast}$, and $(\cdot)^{H}$ denote the transpose, the conjugate, and the conjugate transpose operators, respectively. $\mathbb{E}_{x}\{\cdot\}$ denotes the expectation operator with respect to the variable $x$. The operator, $\hat{x}$ represents the estimated value of $x$. $\Re\{\cdot\}$ and $\Im\{\cdot\}$ denote the real and imaginary parts of a complex quantity. $\mathcal{CN}(\mu,\sigma^2)$ denotes the complex Gaussian distributions with mean $\mu$ and variance $\sigma^2$. Boldface small letters, $\mathbf{x}$ and boldface capital letters, $\mathbf{X}$ are used for vectors matrices, respectively. $[\mathbf{X}]_{x,y}$ represents the entry in row $x$ and column $y$ of $\mathbf{X}$. $\mathbf{I}_{X}$ denotes $X \times X$ identity matrix, $\|\mathbf{x}\|$ represents the $\ell_2$ norm of a vector $\mathbf{x}$, and $\text{diag}(\mathbf{x})$ is used to denote a diagonal matrix, where its diagonal elements are given by the vector $\mathbf{x}$.

\section{System Model} \label{sec:SM}

\begin{figure}[t]
\centering
\vspace{-10pt}
\scalebox{0.95}{\includegraphics {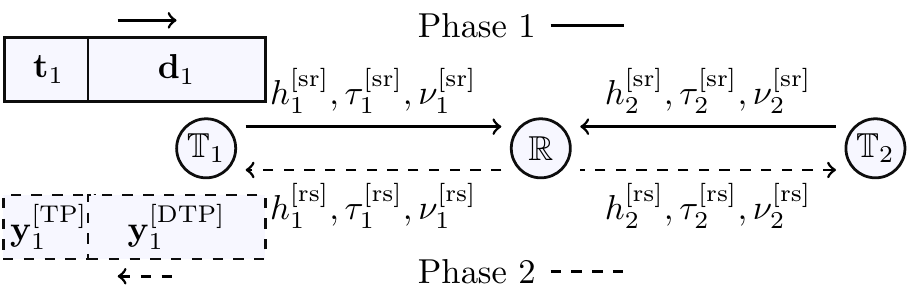}}
   \vspace*{0pt}
   \caption{System Model for AF two-way relay network.}
   \label{fig:BD}
\end{figure}

We consider a half-duplex AF TWRN with two user terminals, $\mathbb{T}_1$ and $\mathbb{T}_2$, and a relay node, $\mathbb{R}$, as shown in Fig. \ref{fig:BD}. All nodes are equipped with a single omnidirectional antenna. The channel gain, timing offset, and carrier frequency offset between the $k$th user terminal and the relay node are denoted by $h_k$, $\tau_k$, and $\nu_k$, respectively, for $k = 1,2$, where the superscripts, $(\cdot)^{\text{[sr]}}$ and $(\cdot)^{\text{[rs]}}$, are used for the parameters from user terminal to relay node and from relay node to user terminal, respectively. The timing and carrier frequency offsets are modeled as \emph{unknown} deterministic parameters over the frame length, which is similar to the approach adopted in \cite{Li-10-A} and \cite{Besson-03-A}. Quasi-static and frequency flat fading channels are considered, i.e., the channel gains do not change over the length of a frame but change from frame to frame according to a complex Gaussian distribution, $\mathcal{CN} (0,\sigma_h^2)$. The use of such channels is motivated by the prior research in this field~\cite{Wang-12-A,Gao-09-A}.

The transmission frame from each user is comprised of training and data symbols. The exchange of data among the two user terminals is completed in two phases:

\begin{enumerate}
\item During the first phase, the transmission frame, $[\mathbf{t}_k,\mathbf{d}_k ]$, is transmitted from the $k$th user, $k = 1,2$, to an intermediate relay node, where $\mathbf{t}_k$ and $\mathbf{d}_k$ denote the $k$th user's  training and data signal, respectively. This is illustrated in Fig. \ref{fig:BD}. The signal from the two users is superimposed at the relay node.

\item During the second phase, the relay node amplifies the superimposed signal and broadcasts it back to the users. The users use the training part of the received signal, $\mathbf{y}_k^\text{[TP]}$, to jointly estimate the \emph{multiple impairments}, i.e, channel gains, timing offsets, and carrier frequency offsets. The effect of these impairments is compensated and the received signal, $\mathbf{y}_k^\text{[DTP]}$, is decoded at the $k$th user's terminal.
\end{enumerate}

Note that the superscripts $(\cdot)^{\text{[TP]}}$ and $(\cdot)^{\text{[DTP]}}$ denote the signals in training and data transmission periods, respectively and Fig. \ref{fig:BD} shows the transmitted frames at the first user terminal, $\mathbb{T}_1$. A similar structure is followed for the second user terminal, $\mathbb{T}_2$.

\newcounter{MYtempeqncnt}

\begin{figure*}[!b]
\normalsize
\setcounter{MYtempeqncnt}{0}
\setcounter{equation}{8} \vspace*{4pt} \hrulefill
\begin{equation}\label{eq:FIM}
\mathbf{F} = \frac{2}{\sigma_u^2}
\left[ {\begin{array}{cccc}
  \Re\{\boldsymbol\Omega^{H} \boldsymbol\Omega\}  &   -\Im\{\boldsymbol\Omega^{H} \boldsymbol\Omega\} &  -\Im\{\boldsymbol\Omega^{H} \mathbf{D} \boldsymbol\Phi \mathbf{t}_2 \} & \Re \{\boldsymbol\Omega^{H} \boldsymbol\Gamma \mathbf{H} \}  \\
  \Im\{\boldsymbol\Omega^{H} \boldsymbol\Omega\}  &   \Re\{\boldsymbol\Omega^{H} \boldsymbol\Omega\} &  \Re\{\boldsymbol\Omega^{H} \mathbf{D} \boldsymbol\Phi \mathbf{t}_2 \} &  \Im \{\boldsymbol\Omega^{H} \boldsymbol\Gamma \mathbf{H} \} \\
  \Im\{\mathbf{t}_2^H \boldsymbol\Phi^H  \mathbf{D} \boldsymbol\Omega \}  & \Re\{\mathbf{t}_2^H \boldsymbol\Phi^H  \mathbf{D} \boldsymbol\Omega \}  & \Re\{\mathbf{t}_2^H \boldsymbol\Phi^H  \mathbf{D}^2 \boldsymbol\Phi \mathbf{t}_2 \} &  \Im\{\mathbf{t}_2^H \boldsymbol\Phi^H \mathbf{D} \boldsymbol\Gamma \mathbf{H} \} \\
 \Re\{\mathbf{H}^{H} \boldsymbol\Gamma^{H} \boldsymbol\Omega \}  & - \Im\{\mathbf{H}^{H} \boldsymbol\Gamma^{H} \boldsymbol\Omega \}  &
 - \Im\{\mathbf{H}^{H} \boldsymbol\Gamma^{H} \mathbf{D} \boldsymbol\Phi^H \mathbf{t}_2 \} &  \Re\{\mathbf{H} \boldsymbol\Gamma^{H} \boldsymbol\Gamma \mathbf{H} \}
\end{array}  } \right]
  \end{equation}
\setcounter{equation}{\value{MYtempeqncnt}}
\end{figure*}

The received signal at the relay node during the training period, $r^\text{[TP]}(t)$, is given by
\begin{align}\label{eq:rt}
r^\text{[TP]}(t) &= \sum_{k=1}^2 h_k^\text{[sr]} e^{j 2 \pi \frac{\nu_k^\text{[sr]}}{T} t} \sum_{n=0}^{L-1} t_k (n) g(t - nT - \tau_k^\text{[sr]} T ) + n(t),
\end{align}
\vspace{-6pt}\\
where the timing and carrier frequency offsets, $\tau_k^\text{[sr]}$ and $\nu_k^\text{[sr]}$, are normalized by the symbol duration $T$, $L$ is the length of training signal $t_k$, $g(t)$ stands for the root-raised cosine pulse function, and $n(t)$ denotes zero-mean complex additive white Gaussian noise (AWGN) at the relay receiver, i.e., $n(t) \sim \mathcal{CN} (0,\sigma_n^{2})$. To avoid amplifier saturation at the relay, the relay node amplifies the received signal, $r^\text{[TP]}(t)$, with the power constraint factor, $\zeta = \frac{1}{\sqrt{2 \sigma_h^{2} + \sigma_n^{2}}}$, and broadcasts the amplified signal to the users \cite{Tian-11-P}. The received signal at the user terminal, $\mathbb{T}_1$, during the training period, $y^\text{[TP]}(t)$ is given by
\begin{align}\label{eq:yt}
y_1^\text{[TP]}(t) = \zeta h_1^\text{[rs]} e^{j 2 \pi \frac{\nu_1^\text{[rs]}}{T} t} \; r^\text{[TP]}(t - \tau_1^\text{[rs]} T) + w_1(t),
\end{align}
where $w_1(t)$ denotes the zero-mean complex AWGN at the receiver of $\mathbb{T}_1$, i.e., $w_1(t) \sim \mathcal{CN} (0,\sigma_w^{2})$. Substituting \eqref{eq:rt} into \eqref{eq:yt}, $y^\text{[TP]}(t)$ is given by
\begin{align}\label{eq:yt2}
y_1^\text{[TP]}(t) &= \zeta h_1^\text{[rs]} e^{j 2 \pi \frac{\nu_1^\text{[rs]}}{T} t} \sum_{k=1}^2 \Big( h_k^\text{[sr]} e^{j 2 \pi \frac{\nu_k^\text{[sr]}}{T} ( t - \tau_1^\text{[rs]} T) )} \notag \\
& \times \sum_{n=0}^{L-1} t_k (n) g(t - nT - \tau_k^\text{[sr]} T - \tau_1^\text{[rs]} T) \Big) \notag \\
&  \hspace{1.2cm} + \zeta h_1^\text{[rs]} e^{j 2 \pi \frac{\nu_1^\text{[rs]}}{T} t} n(t - \tau_1^\text{[rs]} T) + w_1(t)
\end{align}
Note that unlike~\cite{Li-10-A}, the developed system model in \eqref{eq:yt2} takes into account both the timing errors, from users to the relay node, $\tau_k^\text{[sr]}$, $k = 1,2$, and from relay node back to user terminal $\mathbb{T}_1$, $\tau_1^\text{[rs]}$. The received signal in \eqref{eq:yt2}, $y_1^\text{[TP]}(t)$, is sampled with the sampling time $T_s = T/Q$ and the sampled received signal, $y_1^\text{[TP]}(i)$, is given by
\begin{align}\label{eq:yi}
y_1^\text{[TP]}(i) &= \sum_{k=1}^{2} \alpha_k e^{j 2 \pi \nu_k i/Q} \sum_{n=0}^{L-1} t_k (n) g(i T_s - nT - \tau_k T ) \notag \\
&  \hspace{2.2cm} + \zeta h_1^\text{[rs]} e^{j 2 \pi \nu_1^\text{[rs]} i/Q} n(i) + w_1(i)
\end{align}
where
\begin{itemize}
\item $\alpha_k \triangleq \zeta h_k^\text{[sr]} h_1^\text{[rs]} e^{-j 2 \pi \nu_k^\text{[sr]} \tau_1^\text{[sr]} }$ is the combined channel gain from $\mathbb{T}_1$-$\mathbb{R}$-$\mathbb{T}_1$ and $\mathbb{T}_2$-$\mathbb{R}$-$\mathbb{T}_1$ for $k = 1$ and $k=2$, respectively,
\item $\nu_k \triangleq \nu_k^\text{[sr]} +  \nu_1^\text{[rs]}$ is the sum of carrier frequency offsets from $\mathbb{T}_1$-$\mathbb{R}$-$\mathbb{T}_1$ and $\mathbb{T}_2$-$\mathbb{R}$-$\mathbb{T}_1$ for $k = 1$ and $k=2$, respectively, $\nu_1^\text{[sr]} = - \nu_1^\text{[rs]}$ because same oscillators are used during transmission from user $\mathbb{T}_1$ to the relay node and from relay node back to user $\mathbb{T}_1$, thus, $\nu_1 =  \nu_1^\text{[sr]} +  \nu_1^\text{[rs]} = 0$,
\item $\tau_k \triangleq \tau_k^\text{[sr]} +  \tau_1^\text{[rs]}$ is the resultant timing offset from $\mathbb{T}_1$-$\mathbb{R}$-$\mathbb{T}_1$ and $\mathbb{T}_2$-$\mathbb{R}$-$\mathbb{T}_1$ for $k = 1$ and $k=2$, respectively,
\end{itemize}
$Q$ is the sampling factor\footnote{Oversampling is needed to estimate the timing offsets in the presence of pulse shaping.}, $n = 0,1,\hdots,L-1$ and $i = 0,1,\hdots,LQ-1$ are used to denote $T$-spaced and $T_s$-spaced samples, respectively, and $n(i)$ has been used in place of $n(i T_s - \tau_1^\text{[rs]} T)$, since $n(t)$ denotes the AWGN and its statistics are not affected by time delays. Upon reception of signal broadcasted from the relay, it is assumed that the users first employ coarse frame synchronization to ensure that the superimposed signals are within one symbol duration from each other i.e., $\tau_1 - \tau_2 < 1$, and $\tau_1,\tau_2 \in (-0.5,0.5)$. This assumption is in line with prior research in this field \cite{Li-10-A}.

Eq. \eqref{eq:yi} can be written in vector form as
\begin{align}\label{eq:yt1}
\mathbf{y}_1^\text{[TP]} = \alpha_1 \mathbf{G}_1 \mathbf{t}_1 +  \alpha_2 \boldsymbol\Lambda_2 \mathbf{G}_2 \mathbf{t}_2 + \zeta h_1^\text{[rs]} \boldsymbol\Lambda^\text{[rs]} \mathbf{n} + \mathbf{w}_1
\end{align}
where
\begin{itemize}
\item $\mathbf{G}_k$ is the $LQ \times L$ matrix of the samples of the pulse shaping filter such that $\left[\mathbf{G}_k \right]_{i,n} \triangleq g_\text{rrc}(iT_s-nT-\tau_k^{\text{[rd]}}T)$,
\item $\boldsymbol\Lambda_2 \triangleq \text{diag} \left( [e^{j 2 \pi \nu_2 (0) /N},\hdots,e^{j 2 \pi \nu_2 (LQ-1)/N}] \right)$ is an $LQ \times LQ$ matrix,
\item  $\boldsymbol\Lambda^\text{[rs]} \triangleq \text{diag} \left( [e^{j 2 \pi \nu^\text{[rs]} (0) /N},\hdots,e^{j 2 \pi \nu^\text{[rs]} (LQ-1)/N}] \right)$ is an $LQ \times LQ$ matrix
\item  $ \mathbf{y}_1^\text{[TP]} \triangleq [ y_1^\text{[TP]}(0),\hdots,y_1^\text{[TP]}(LQ-1)]^T$
\item $ \mathbf{t}_k \triangleq [ t_k(0),\hdots,t_k(L-1) ]^T$,
\item $ \mathbf{n}\triangleq [ n(0),\hdots,n(LQ-1) ]^T$, and
\item $ \mathbf{w}_1\triangleq [ w_1(0),\hdots,w_1(LQ-1) ]^T$.
\end{itemize}
The received signal during the data transmission period, $\mathbf{y}_1^\text{[DTP]}$, can be similarly expressed as \eqref{eq:yt1}, where training $\mathbf{t}_k$ is replaced by the data $\mathbf{d}_k \triangleq [ d_k(0),\hdots,d_k(L-1) ]^T$. Note that as anticipated, the data length $L$ is different and larger than the training length $L$ as discussed in Section \ref{sec:SR}.

Next, without loss in generality, we derive the CRLB and estimators for joint estimation of channel gains, timing offsets, and carrier frequency offsets at the user terminal $\mathbb{T}_1$. Note that the system model in this section and the derived CRLB, estimation, and detection schemes in the following sections can be easily manipulated to detect $\mathbf{d}_1$ at the user terminal $\mathbb{T}_2$. These details are not included to avoid repetition.

\section{Cram\'{e}r-Rao Lower Bound} \label{sec:CRLB}

In this section, the CRLB for joint estimation of multiple impairments at $\mathbb{T}_1$ are derived. The signal model in \eqref{eq:yt1} can be rewritten as
\begin{align}\label{eq:y2}
\mathbf{y}_1^\text{[TP]} = \boldsymbol\Omega \boldsymbol\alpha + \mathbf{u},
\end{align}
where $ \boldsymbol\Omega \triangleq [ \mathbf{G}_1 \mathbf{t}_1 \; \boldsymbol\Lambda_2 \mathbf{G}_2 \mathbf{t}_2 ]$ is an $LQ \times 2$ matrix, $\boldsymbol\alpha \triangleq [ \alpha_1,\alpha_2]^T$, and $ \mathbf{u} \triangleq \zeta h_1^\text{[rs]} \boldsymbol\Lambda^\text{[rs]} \mathbf{n} + \mathbf{w}_1$. Based on the assumptions and proposed system model in Section \ref{sec:SM}, the received signal vector, $\mathbf{y}_1^\text{[TP]}$,
is a circularly symmetric complex Gaussian random variable,
$\mathbf{y}^\text{[TP]} \sim \mathcal{CN} (\boldsymbol\mu,\boldsymbol\Sigma)$,
with mean $\boldsymbol\mu$ and covariance matrix $\boldsymbol\Sigma$, given by
\begin{subequations}
\begin{align}
\boldsymbol\mu &= \boldsymbol\Omega  \boldsymbol\alpha,  \, \, \, \, \text{and}\\
\boldsymbol\Sigma &= \mathbb{E} \{ \mathbf{u} \mathbf{u}^H \} = ( \zeta^2 \sigma_h^2 \sigma_n^2 + \sigma_w^2) \mathbf{I}_{LQ} = \sigma_u^2 \mathbf{I}_{LQ},
\end{align}
\end{subequations}
respectively. To determine the CRLB, we have to first formulate the parameter vector of interest. The user $\mathbb{T}_1$ has to estimate the channel gains, $\boldsymbol\alpha$, timing offsets $\boldsymbol\tau \triangleq [\tau_1,\tau_2]^T$, and the carrier frequency offset $\nu_2$. There is no need to estimate $\nu_1$ as this is found to be $0$ as explained below \eqref{eq:yi}. As a result, the parameter vector of interest, $\boldsymbol\lambda$, is given by
\begin{align}
 \boldsymbol\lambda \triangleq [ \Re\{\boldsymbol\alpha\}^{T}, \Im\{\boldsymbol\alpha\}^{T} , \nu_2 , \boldsymbol\tau^{T} ]^{T}
 \end{align}
In the following, we derive Fisher's information matrix (FIM) for the estimation of $\boldsymbol\lambda$.

\emph{Theorem 1:} Based on the proposed system model, the FIM, denoted by $\mathbf{F}$, for the estimation of $\boldsymbol\lambda$ is given by \eqref{eq:FIM}, at the bottom of this page, where
\begin{itemize}
\item $\boldsymbol\Gamma \triangleq [ \mathbf{R}_1 \mathbf{t}_1 \; \boldsymbol\Lambda_2 \mathbf{R}_2 \mathbf{t}_2 ]$ is an $LQ \times 2$ matrix, $\mathbf{R}_k \triangleq \frac{\partial \mathbf{G}_k}{\partial \tau_k}$ is an $LQ \times 2$ matrix,
\item $\mathbf{D}  \triangleq 2 \pi \times \text{diag} \{0,\hdots,LQ-1\}$ is an $LQ \times LQ$ matrix,
\item $ \boldsymbol\Phi \triangleq \alpha_2 \boldsymbol\Lambda_2 \mathbf{G}_2 $ is an $LQ \times L$ matrix, and
\item $\mathbf{H} \triangleq  \text{diag} \{ \alpha_1,\alpha_2 \}$ is a $2 \times 2$ matrix.
\end{itemize}

\emph{Proof:} The FIM is derived by following the steps given in \cite{Kay-93-B}. The detailed derivation is omitted due to space limitation.

Finally, the CRLB for the estimation of $\boldsymbol\lambda$ is given by the diagonal elements of the inverse of $\mathbf{F}$. Note that the CRLB for channel estimation is the sum of the CRLBs for real and imaginary parts of the channel estimation~\cite{Kay-93-B}.
\addtocounter{equation}{1}

\section{Joint Parameter Estimation} \label{sec:EST}

In this section, the ML estimator for joint estimation of multiple impairments in AF TWRN is derived. Subsequently, the DE based estimator is applied to reduce the computational complexity for obtaining these impairments.

\subsection{ML Estimation}\label{sec:ML_EST}

Based on the signal model in \eqref{eq:y2}, the ML estimates of the parameters, $\boldsymbol\alpha, \boldsymbol\tau$, and $\nu_2$, can be determined by minimizing the cost function, $\mathbf{J}$, according to \vspace{-0pt}
\begin{align}\label{eq:JSAF}
\mathbf{J}(\boldsymbol\alpha,\boldsymbol\tau,\nu_2)
= \left\| \mathbf{y}_1^\text{[TP]} -  \boldsymbol\Omega \boldsymbol\alpha \right\|^2.
\end{align}
Given $\boldsymbol\tau$ and $\nu_2$, it is straightforward to show that the ML estimate of $\boldsymbol\alpha$, denoted by $\hat{\boldsymbol\alpha}$, can be determined as
\begin{equation}\label{eq:alpha_cap_SAF}
\hat{\boldsymbol\alpha} = \left(\boldsymbol\Omega^{H} \boldsymbol\Omega\right)^{-1} \boldsymbol\Omega^{H} \mathbf{y}_1^\text{[TP]}.
\end{equation}
Substituting~(\ref{eq:alpha_cap_SAF}) in~(\ref{eq:JSAF}), the estimates of $\tau_1,\tau_2$, and $\nu_2$ are obtained via
\begin{equation}\label{eq:maxSAF}
\hat{\tau}_1, \hat{\tau}_2, \hat{\nu}_2 = \arg
\min_{\tau_1,\tau_2,\nu_2} \underbrace{-\left(\mathbf{y}_1^\text{[TP]}\right)^{H}
\boldsymbol\Omega \left(\boldsymbol\Omega^{H} \boldsymbol\Omega\right)^{-1}
\boldsymbol\Omega^{H} \mathbf{y}_1^\text{[TP]}}_{\triangleq\chi\left(\tau_1,\tau_2,\nu_2\right)},
\end{equation}
where $\arg \min$ denotes the arguments, $\tau_1,\tau_2$, and $\nu_2$, that minimize the expression $\chi\left(\tau_1,\tau_2,\nu_2\right)$ and $\mathbf{y}_1^\text{[TP]}$ is defined in \eqref{eq:y2}. The channel estimates, $\hat{\alpha}_1$ and $\hat{\alpha}_2$, are obtained by substituting $\hat{\tau}_1, \hat{\tau}_2$, and $\hat{\nu}_2$ back into~(\ref{eq:alpha_cap_SAF}).

The minimization in~\eqref{eq:maxSAF} requires a 3-dimensional exhaustive search over the discretized set of possible timing and frequency offset values, which is inherently very computationally complex. Furthermore, to reach the CRLB (see Fig.~\ref{fig:mse} in Section~\ref{sec:SR}), the exhaustive search in (\ref{eq:maxSAF}) needs to be carried out with a very high resolution\footnote{Step sizes of $10^{-2}$ and $10^{-4}$ for MTOs and MCFOs, respectively.}, which further increases the complexity of the proposed ML estimator. In the following subsection, DE is employed as a computationally efficient algorithm to carry out the minimization in \eqref{eq:maxSAF} \cite{Price-05-B}.

\subsection{Differential Evolution based Estimation}\label{sec:DE}

DE and genetic algorithms are considered as a subclass of \emph{evolutionary algorithms} since they attempt to evolve the solution for a problem through recombination, mutation, and survival of the fittest. More specifically, DE is an optimization algorithm, where a number of parameter vectors are generated and updated at each iteration in order to reach the solution \cite{Price-05-B}. Following the detailed steps and parameterization of the DE algorithm outlined in \cite{Nasir-13-A-TSP} and changing the estimation vector length to $3$ parameters, $\hat{\tau}_1, \hat{\tau}_2$, and $\hat{\nu}_2$, the minimization in \eqref{eq:maxSAF} can proceed. Substituting these estimates in (\ref{eq:alpha_cap_SAF}) also generates the desired channel estimates.

\newtheorem{remark}{Remark}
\begin{remark}\label{rem:5}
The computational requirements of the ML and the DE algorithms are quantified using CPU execution time~\cite{Moller-08-A}. The execution time is observed by setting training length $L = 80$, when an Intel Core i7-2670QM CPU @ 2.20 GHz processor with 8 GB of RAM is used. It has been observed that comparing to the ML estimator, the DE algorithm is capable of estimating the multiple impairments approximately $10^4$ times more quickly.

The computational complexity of the ML and the DE algorithms can also be compared by calculating the number of additions plus multiplications. By following the steps outlined in \cite[(20)-(21)]{Nasir-13-A-TSP}, we find that an ML algorithm requires $1.84 \times 10^{13}$ multiplications and additions, however the DE algorithm needs $1.66 \times  10^{9}$ multiplications and additions in order to estimate the multiple impairments. This method of computational complexity comparison also verifies that DE algorithm is capable of estimating the multiple impairments approximately $10^4$ times more quickly.

Note that a large number of additions and multiplications are not the point of concern here. This is because the proposed ML and DE estimation methods are applied for initialization only once at system start-up. Afterwards, the estimates of previously transmitted frames may be used to update the new estimates since timing and carrier frequency offsets do not rapidly change from frame to frame. This is due to the fact that oscillator properties are mainly affected by temperature and other physical phenomena that do not rapidly fluctuate with time~\cite{Singh-06-B}.
\end{remark}

\section{MMSE Receiver and Data Detection} \label{sec:SD}

\begin{figure*}[t]
    \centering
    \begin{minipage}[h]{0.48\textwidth}
    \centering
    \includegraphics[width=1 \textwidth]{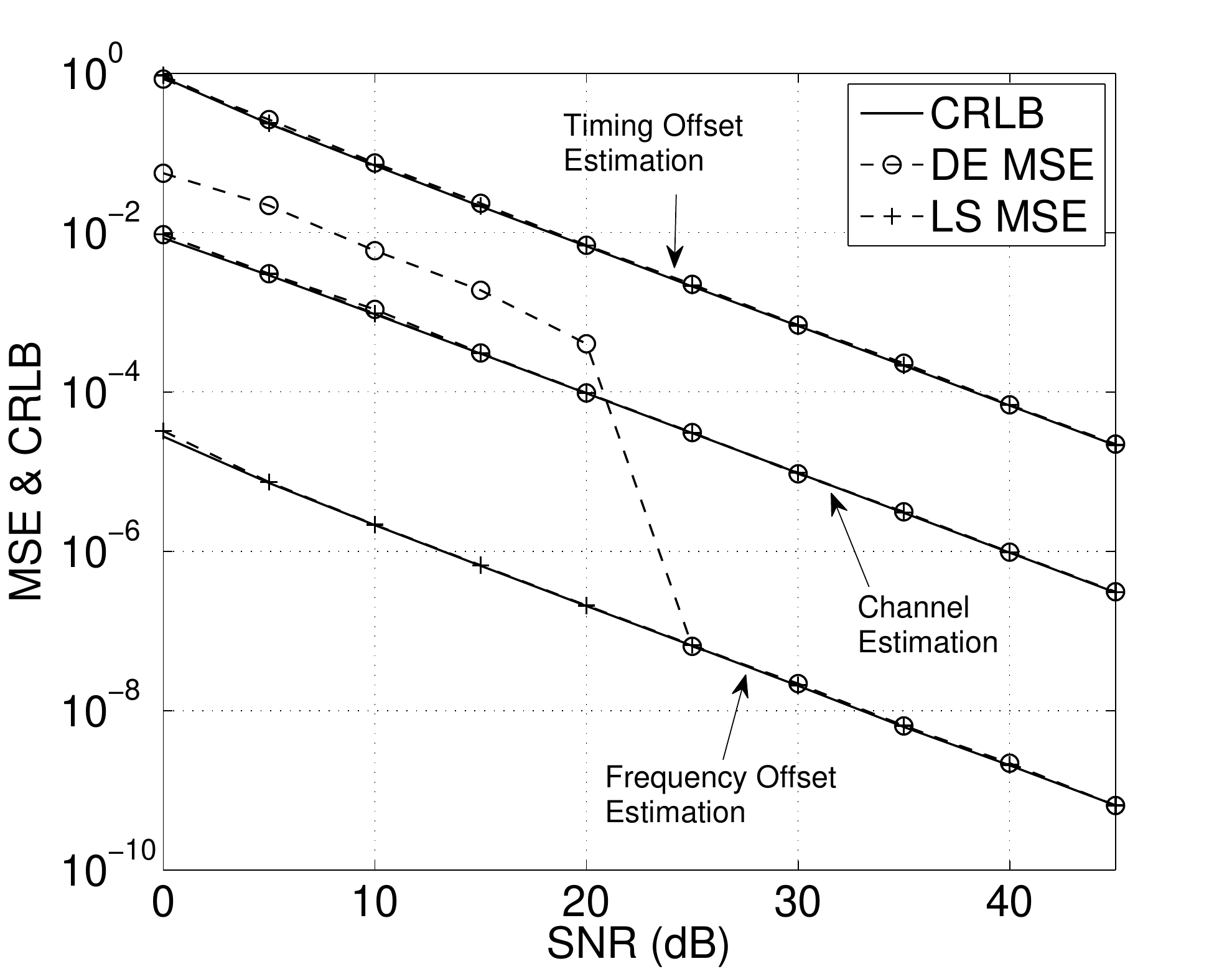}
  \caption{CRLBs and MSE for ML $\&$ DE based estimation of channel gains, timing offsets, and carrier frequency offsets.}
  \label{fig:mse}
  \end{minipage}
    \hspace{0.3cm}
    \begin{minipage}[h]{0.48\textwidth}
    \centering
    \includegraphics[width=1 \textwidth]{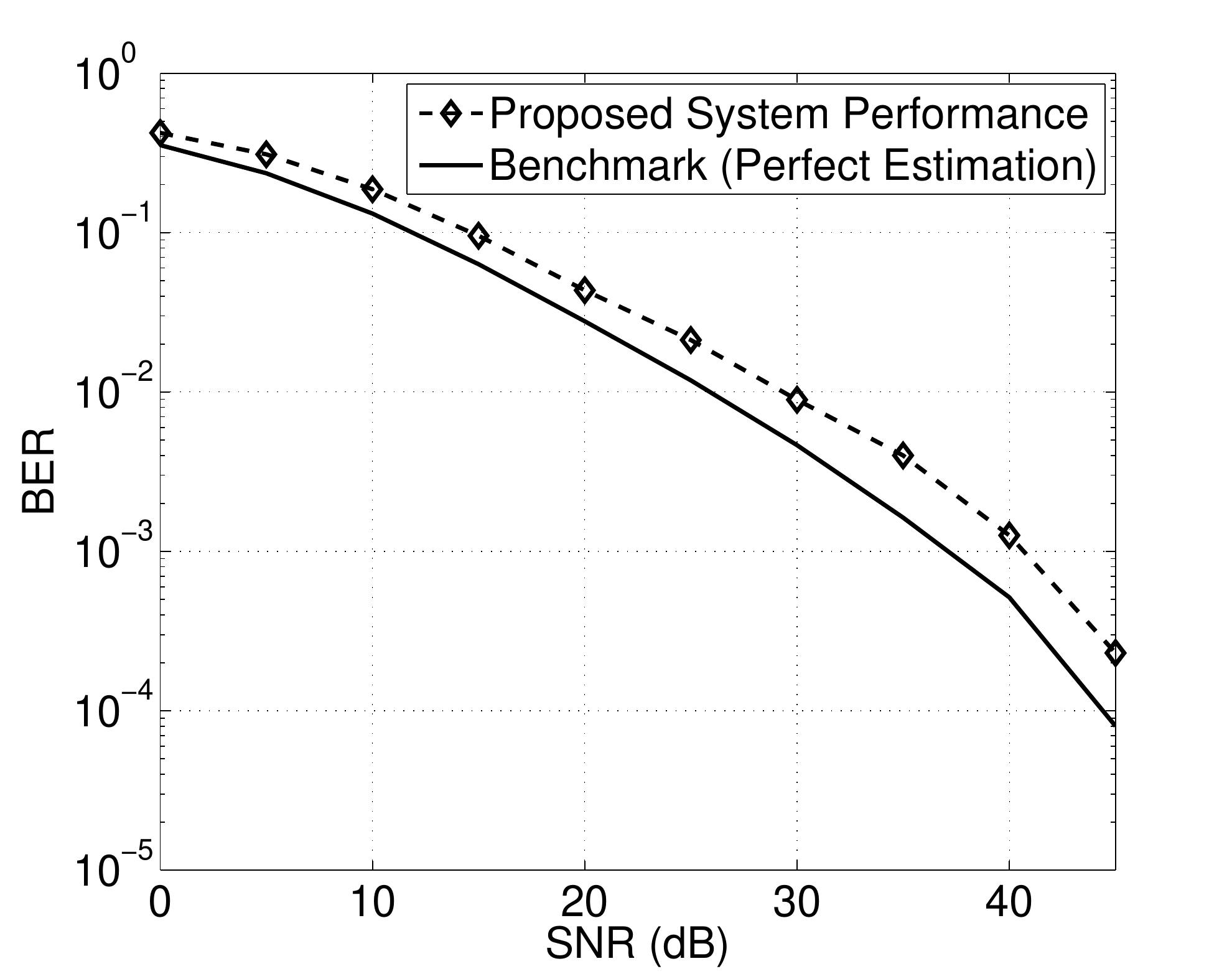}
  \caption{BER of the proposed AF TWRN with DE based estimation and perfect estimation.}
  \label{fig:ber}
  \end{minipage}
  \vspace{-0.15in}
\end{figure*}

In this section, an MMSE receiver for compensating the effect of impairments and detecting the signal from user $\mathbb{T}_2$ is derived. Following \eqref{eq:yt1}, the received signal at user $\mathbb{T}_1$ during the data transmission period, $\mathbf{y}_1^\text{[DTP]}$, is given by
\begin{align}\label{eq:yd1}
\mathbf{y}_1^\text{[DTP]} = \alpha_1 \mathbf{G}_1 \mathbf{d}_1 +  \alpha_2 \boldsymbol\Lambda_2 \mathbf{G}_2 \mathbf{d}_2 + \zeta h_1^\text{[rs]} \boldsymbol\Lambda^\text{[rs]} \mathbf{n} + \mathbf{w}_1
\end{align}
The user $\mathbb{T}_1$ has to decode the signal $\mathbf{d}_2$ using the received signal $\mathbf{y}_1^\text{[DTP]}$, the estimated impairments, $\hat{\boldsymbol\alpha}$, $\hat{\boldsymbol\tau}$, $\hat{\nu_2}$, and its own data $\mathbf{d}_1$. Let us define an $LQ \times 1$ vector $\mathbf{z} \triangleq \mathbf{y}_1^\text{[DTP]} - \alpha_1 \mathbf{G}_1 \mathbf{d}_1$. Using $\mathbf{y}_1^\text{[DTP]}$, $\mathbf{d}_1$, $\hat{\tau}_1$, $\hat{\alpha}_1$, user $\mathbb{T}_1$ can estimate the vector $\mathbf{z}$ as
\begin{align}
\hat{\mathbf{z}} &\triangleq \mathbf{y}_1^\text{[DTP]} - \hat{\alpha}_1 \hat{\mathbf{G}}_1 \mathbf{d}_1 \notag \\
&= \alpha_2 \boldsymbol\Lambda_2 \mathbf{G}_2 \mathbf{d}_2 + \mathbf{u}
\end{align}
where $\hat{\mathbf{G}}_1 = \mathbf{G}_1 |_{\tau_1 = \hat{\tau}_1}$ and $\mathbf{u} \triangleq \zeta h_1^\text{[rs]} \boldsymbol\Lambda^\text{[rs]} \mathbf{n} + \mathbf{w}_1$ is defined in \eqref{eq:y2}. Applying MMSE based detection, the signal from user $\mathbb{T}_2$ can be decoded as
\begin{align}\label{eq:d2}
\hat{\mathbf{d}}_2 = ( \hat{\boldsymbol\Phi}^H \hat{\boldsymbol\Phi} + \sigma_u^2 \mathbf{I}_L ) \hat{\boldsymbol\Phi}^H  \hat{\mathbf{z}}
\end{align}
where $ \hat{\boldsymbol\Phi} \triangleq \hat{\alpha}_2 \hat{\boldsymbol\Lambda}_2 \hat{\mathbf{G}}_2 $, $\hat{\boldsymbol\Lambda}_2 = {\boldsymbol\Lambda}_2 |_{\nu_2 = \hat{\nu}_2}$, and $\hat{\mathbf{G}}_2 = \mathbf{G}_2 |_{\tau_2 = \hat{\tau}_2}$. In order to \emph{benchmark} the decoding performance of the overall AF TWRN, the benchmark data detection $\hat{\mathbf{d}}^\text{[BM]}_2$, that is based on the perfect knowledge of multiple impairments, is given by
\begin{align}\label{eq:d2BM}
\hat{\mathbf{d}}^\text{[BM]}_2 = ( {\boldsymbol\Phi}^H {\boldsymbol\Phi} + \sigma_u^2 \mathbf{I}_L ) {\boldsymbol\Phi}^H  \hat{\mathbf{z}}
\end{align}
where $ {\boldsymbol\Phi} \triangleq {\alpha}_2 {\boldsymbol\Lambda}_2 {\mathbf{G}}_2 $.

\vspace{-5pt}

\section{Simulation Results} \label{sec:SR}

In this section, we present simulation results to evaluate the estimation and BER performance of the AF TWRN. The training length is set to $L = 80$ and data transmission length is set to $L = 400$ symbols, which results in the synchronization overhead of $16.6 \%$. Without loss of generality, it is assumed that during the training period, linearly independent, unit-amplitude phase shift keying (PSK) training signals are transmitted from two users. Such TSs have also been considered previously, e.g., in \cite{Li-10-A}.\footnote{The design of the optimal training sequences is outside the scope of this paper.\vspace{-0pt}}. During the data transmission period, quadrature phase-shift keying (QPSK) modulation is employed for data transmission. The oversampling factor is set to $Q=2$ and the noise variances, $\sigma_n^2 = \sigma_w^2 = 1/\text{SNR}$. The synchronization parameters, $\tau_1$, $\tau_2$, and $\nu_2$, are assumed to be uniformly distributed over the range $(-0.5,0.5)$. All the channel gains are modeled as independent and identically distributed complex Gaussian random variables with $\mathcal{CN}(0,1)$. In the following, the MSE and BER simulation results are averaged out over $600$ frames with $400$ data symbols per frame, where random realization of Rayleigh fading channel gains is generated every frame.

Fig. \ref{fig:mse} plots the CRLBs, derived in Section \ref{sec:CRLB}, and estimation MSE for joint estimation of multiple impairments. Without loss of generality, the CRLBs and the MSE estimation performance for $\alpha_1$, $\tau_1$, and $\nu_2$ are presented only, where similar results to that of $\alpha_1$ and $\tau_1$ are observed for $\alpha_2$ and $\tau_2$, respectively. Fig. \ref{fig:mse} shows that the mean-square estimation error of the proposed ML and DE estimators is very close to the derived CRLBs at moderate-to-high SNRs. Note that the mean square estimation error of ML estimator is close to the derived CRLBs for whole considered range of SNR (0-45 dB). On the other hand, for the DE based estimation, the MSE for frequency offset estimation approaches to the CRLB after $25$ dB. This is because, unlike the ML estimator, the DE estimator is not based on the exhaustive search criteria and is more computationally efficient than ML estimator.

Fig. \ref{fig:ber} illustrates the BER performance of overall AF TWRN, that employs the DE based estimator and MMSE receiver to decode the received signal. The DE based estimator is employed because ML estimator is computationally very complex (see Remark \ref{rem:5} in Section \ref{sec:EST}). The BER performance of the proposed estimation and decoding schemes ($\hat{\mathbf{d}}_2$ in \eqref{eq:d2}) is compared with the benchmark decoding scheme ($\hat{\mathbf{d}}^\text{[BM]}_2$ in \eqref{eq:d2BM}), which assumes perfect knowledge of multiple impairments. Fig. \ref{fig:ber} shows that the BER performance of the proposed overall AF TWRN is close to the performance of a TWRN that is based on the assumption of perfect knowledge of synchronization parameters, i.e., there is a performance gap of just $2$-$3$ dB only at moderate-to-high SNRs. To the best of our knowledge, our algorithm is the first complete solution for the joint estimation and compensation of channel gains, timing offsets, and carrier frequency offsets. Hence, the performance of the proposed algorithm cannot be compared with any existing algorithm. For example, the algorithm in \cite{Gao-09-A}, which estimates and compensates the effect of channel parameters in AF TWRN, fails to decode the received signal and shows very poor BER performance in the presence of multiple synchronization impairments.

\vspace{-3pt}

\section{Conclusions} \label{sec:CON}

This paper has proposed a system model for achieving AF TWRN synchronization in the presence of multiple impairments, i.e., channel gains, timing offsets, and carrier frequency offsets. In order to extract the user's information from the received signal, each user jointly estimates the multiple impairments and compensates for their effect from the received signal. CRLBs for joint estimation of multiple impairments are derived and simulation results show that the mean-square estimation error of the applied ML and DE estimators is very close to the derived CRLBs. Next, MMSE based received is derived to compensate for the effect of multiple impairments and decode the user's information. The BER performance of overall AF TWRN, employing the proposed estimation and decoding schemes, is close to the lower bound BER (performance gap of $2$-$3$ dB only), that assumes perfect knowledge of multiple impairments.

\vspace{-4pt}

\balance


\begin{thebibliography}{10}
\providecommand{\url}[1]{#1}
\csname url@samestyle\endcsname
\providecommand{\newblock}{\relax}
\providecommand{\bibinfo}[2]{#2}
\providecommand{\BIBentrySTDinterwordspacing}{\spaceskip=0pt\relax}
\providecommand{\BIBentryALTinterwordstretchfactor}{4}
\providecommand{\BIBentryALTinterwordspacing}{\spaceskip=\fontdimen2\font plus
\BIBentryALTinterwordstretchfactor\fontdimen3\font minus
  \fontdimen4\font\relax}
\providecommand{\BIBforeignlanguage}[2]{{%
\expandafter\ifx\csname l@#1\endcsname\relax
\typeout{** WARNING: IEEEtran.bst: No hyphenation pattern has been}%
\typeout{** loaded for the language `#1'. Using the pattern for}%
\typeout{** the default language instead.}%
\else
\language=\csname l@#1\endcsname
\fi
#2}}
\providecommand{\BIBdecl}{\relax}
\BIBdecl

\bibitem{Laneman-A-03}
J.~N. Laneman and G.~W. Wornell, ``Distributed space-time-coded protocols for
  exploiting cooperative diversity in wireless networks,'' \emph{{IEEE} Trans.
  Inf. Theory}, vol.~49, no.~10, pp. 2415--2425, Oct. 2003.

\bibitem{Oechtering-08-A}
T.~J. Oechtering, C.~Schnurr, I.~Bjelakovic, and H.~Boche, ``Broadcast capacity
  region of two-phase bidirectional relaying,'' \emph{{IEEE} Trans. Inf.
  Theory}, vol.~54, no.~1, pp. 454--458, Jan. 2008.

\bibitem{Cheng-12-A}
P.~Cheng, L.~Gui, Y.~Rui, Y.~J. Guo, X.~Huang, and W.~Zhang, ``Compressed
  sensing based channel estimation for two-way relay networks,'' \emph{IEEE
  Wireless Communication Letters}, vol.~1, no.~3, pp. 201--204, Jun. 2012.

\bibitem{Tian-11-P}
S.~Tian, Y.~Li, and B.~Vucetic, ``A near optimal amplify and forward relaying
  in two-way relay networks,'' in \emph{Proc. IEEE ICC}, 2011.

\bibitem{Tian-10-A}
Y.~Tian, X.~Lei, Y.~Xiao, and S.~Li, ``{SAGE} based joint timing-frequency
  offsets and channel estimation in distributed {MIMO} systems,''
  \emph{Elsevier Journal Computer Commun.}, vol.~33, no.~17, pp. 2125--2131,
  Jul. 2010.

\bibitem{Nasir-13-A-TSP}
A.~A. Nasir, H.~Mehrpouyan, S.~Durrani, S.~D. Blostein, R.~A. Kennedy, and
  B.~Ottersten, ``Transceiver design for distributed {STBC} based {AF}
  cooperative networks in the presence of timing and frequency offsets,''
  \emph{{IEEE} Trans. Signal Process.}, vol.~61, no.~12, pp. 3143--3158, Jun.
  2013.

\bibitem{Abdallah-12-A}
S.~Abdallah and I.~N. Psaromiligkos, ``Blind channel estimation for
  amplify-and-forward two-way relay networks employing {M-PSK} modulation,''
  \emph{{IEEE} Trans. Signal Process.}, vol.~60, no.~7, pp. 3604--3615, Jul.
  2012.

\bibitem{Abdallah-13-A}
------, ``{EM}-based semi-blind channel estimation in amplify-and-forward
  two-way relay networks,'' \emph{IEEE Wireless Communication Letters}, vol.~2,
  no.~5, pp. 2162--2337, Oct. 2013.

\bibitem{Jiang-13-A}
Z.~Jiang, H.~Wang, and Z.~Ding, ``A bayesian algorithm for joint symbol timing
  synchronization and channel estimation in two-way relay networks,''
  \emph{{IEEE} Trans. Commun.}, vol.~61, no.~10, pp. 4271--4283, Oct. 2013.

\bibitem{Li-10-A}
X.~Li, C.~Xing, Y.-C. Wu, and S.~C. Chan, ``Timing estimation and
  resynchronization for amplify-and-forward communication systems,''
  \emph{{IEEE} Trans. Signal Process.}, vol.~58, no.~4, pp. 2218--2229, Apr.
  2010.

\bibitem{Wang-12-A}
C.~Wang, T.~C.-K. Liu, and X.~Dong, ``Impact of channel estimation error on the
  performance of amplify-and-forward two-way relaying,'' \emph{{IEEE} Trans.
  Veh. Technol.}, vol.~61, no.~3, pp. 1197--1207, Mar. 2012.

\bibitem{Gao-09-A}
F.~Gao, C.~Schnurr, I.~Bjelakovic, and H.~Boche, ``Optimal channel estimation
  and training design for two-way relay networks,'' \emph{{IEEE} Trans.
  Commun.}, vol.~57, no.~10, pp. 3024--3033, Oct. 2009.

\bibitem{Wang-10-P}
G.~Wang, F.~Gao, Y.-C. Wu, and C.~Tellambura, ``Joint {CFO} and channel
  estimation for {CP}-{OFDM} modulated two-way relay networks,'' in \emph{Proc.
  IEEE ICC}, 2010.

\bibitem{Price-05-B}
K.~V. Price, R.~M. Storn, and J.~A. Lampinen, \emph{Differential {E}volution:
  {A} practical approach to global optimization.}, G.~Rozenberg, T.~Bäck,
  J.~Kok, H.~Spaink, and A.~E. Eiben, Eds.\hskip 1em plus 0.5em minus
  0.4em\relax Springer-Verlag Berlin Heidelberg, 2005.

\bibitem{Besson-03-A}
O.~Besson and P.~Stoica, ``On parameter estimation of {MIMO} flat-fading
  channels with frequency offsets,'' \emph{{IEEE} Trans. Signal Process.},
  vol.~51, no.~3, pp. 602--613, Mar. 2003.

\bibitem{Kay-93-B}
S.~M. Kay, \emph{Fundamentals of Statistical Signal Processing: Estimation
  Theory}.\hskip 1em plus 0.5em minus 0.4em\relax NJ: Prentice Hall, 1993.

\bibitem{Moller-08-A}
N.~Moller, ``On {S}choonhage's algorithm and subquadratic integer gcd
  computation,'' \emph{Mathematics of Computation}, vol.~77, pp. 589--607, Jan.
  2008.

\bibitem{Singh-06-B}
B.~P. Singh and R.~Singh, \emph{Electronic devices and integrated
  circuits}.\hskip 1em plus 0.5em minus 0.4em\relax Pearson Education India,
  2006.

\end{thebibliography}
\end{document}